\documentclass{desyproc}
\usepackage{hyperref}

\begin{document}

\newcommand{\G}{\gamma}
\newcommand{\GP}{\gamma^\prime}
\newcommand{\too}{\rightarrow}
\newcommand{\mix}{\chi}
\newcommand{\mixo}{\chi_{_0}}
\newcommand{\OP}{\omega_\mathrm{P}}
\newcommand{\OPPP}{\omega_\mathrm{P}^{2\prime}}
\newcommand{\muu}{ m_{\GP}} 
\newcommand{\BSP}{\hspace{1cm};\hspace{1cm}}
\newcommand{\OPc}{\omega_{\mathrm{P}\odot}}
\newcommand{\MV}{ v_{_\mathrm{M\ddot{o}l}    } }
\newcommand{\rmi}{\mathrm{i}}
\newcommand{\eff}{{\rm eff}}
\renewcommand\({\left(}
\renewcommand\){\right)}
\renewcommand\[{\left[}
\renewcommand\]{\right]}
\newcommand{\pa}{\partial}
\newcommand{\dd}{{\rm d}}
\newcommand{\e}{{\rm e}}
\def\be{\begin{equation}}
\def\ee{\end{equation}}
\def\bea{\begin{eqnarray}}
\def\eea{\end{eqnarray}}
\newcommand{\gs}{\delta_{_{\rm GS}}}
\newcommand\vp{\varphi}
\newcommand\eps{\epsilon}
\newcommand\mpl{m_{\rm p}}
\newcommand{\mcL}{{\mathcal L}}
\newcommand{\1}{\mathbbm{1}}
\newcommand{\GeV}{{\rm GeV}}
\newcommand{\MeV}{{\rm MeV}}
\newcommand{\keV}{{\rm keV}}
\newcommand{\abs}{\Gamma_{\rm abs}}

\title{
\vspace{-2cm}
\hfill{\small DESY 08-130}\\[1cm]
Bounds on Very Weakly Interacting Sub-eV Particles (WISPs) from Cosmology and Astrophysics}

\author{{\slshape Javier Redondo}\\[1ex]
$^1$DESY theory group, Notkestra{\ss}e 85, 22607 Hamburg, Germany}

\contribID{redondo\_javier}

\desyproc{DESY-PROC-2008-02}
\acronym{Patras 2008} 
\doi  

\maketitle

\begin{abstract}
Many weakly interacting sub-electronVolt particles (WISPs) are easily accommodated in extensions of the standard model. Generally the strongest bounds on their existence come from stellar evolution and cosmology, where to the best of our knowledge observations seem to agree with the standard budget of particles. In this talk I review the most demanding constraints for axions and axion-like-particles, hidden photons and  mini-charged particles.
\end{abstract}
\vspace{.6cm}
\noindent
There is little doubt in the particle physics community about the need of complementing the already very successful standard model (SM) to pursue a completely satisfactory final theory of elementary particles. 
On the other hand, and with the exception of the dark matter, our increasingly precise knowledge of the universe shows no trace of physics beyond the SM. 
If new light particles exist they should be very weakly interacting, probably only accessible to extremely precise experiments. Experiments such as the ones presented in this conference.  

Astrophysics and cosmology are often strong probes of weakly interacting particles. The reason is clear: the huge magnitudes of the typical sizes, time scales, densities or temperatures in the early universe or in stars can convert a tiny ``microscopic" effect in a big qualitative change in the evolution of the whole system. This conclusion is specially emphasized when we note that the only weakly interacting sub-eV particles (WISPs) in the standard model are neutrinos, whose production cross sections are strongly energy-dependent and therefore their role is increasingly inhibited as temperatures drop below the electroweak scale. 
Thus, in an non-extreme range of temperatures the early universe and stellar plasmas are very opaque to standard particles and WISPs can be the most efficient way of energy transfer. Whenever such an anomalous energy transfer has an observable implication we can derive strong constraints on the WISP interactions with the standard particles constituting the relevant plasma.

The oldest picture of the universe we have is a dense and hot plasma of elementary particles that expanded against gravity. As this plasma cooled down, the three long range forces clustered the particles into the structures which nowadays are found: the color force first confined quarks into protons and neutrons and later merged them into light nuclei (at BBN), the Coulomb force combined them with electrons into atoms (releasing the CMB) that gravity finally clustered into galaxies, then into clusters, etc...
After the first galaxies formed, the conditions for stars to be born were settled.
During all these steps of structure formation (in a broad sense) the role of WISPs can be constrained. 
Let us start this review in chronological order. Summary plots on the reviewed bounds are shown at the end of this contribution.

\emph{Big Bang Nucleosynthesis.-}
BBN left an invaluable probe of the early universe environment imprinted in today's observable light nuclei abundances \cite{Iocco:2008va}.
Below $T\sim 0.7$ MeV the weak reactions $p+e^-\leftrightarrow n+ \nu_e$ became ineffective, fixing the neutron/proton density ratio to $n/p\sim 1/7$. 
All particles present contribute to the energy density $\rho$ which determines the speed of the cosmic expansion $H\propto \sqrt{\rho}$ and the ``freeze-out" ratio $n/p$ in turn. The larger $H$ the sooner the $p$-$n$ freezing and the higher $n/p$.
Later, all neutrons are confined into $^4$He nuclei whose primordial abundance can be measured today, leading to a bound on the non-standard energy density $\rho_x$ during BBN, usually expressed as an effective number of thermal neutrino species, 
$N_{\nu,x}^{\rm eff}\equiv  \frac{4}{7}\frac{30}{\pi^2 T^4}\rho_x= -0.6_{-0.8}^{+0.9}$ \cite{Simha:2008zj}, where we assumed three standard neutrinos.

Therefore, while a spin-zero particle thermalized during BBN is allowed, this is not the case for other\footnote{For details of these hypothetical particles and their embedding in theories beyond the SM the reader is refereed to the contributions of Andreas Ringwald and Joerg Jaeckel in these proceedings.} WISPs like a mini-charged particle (MCP) ($N_{\nu,\rm MCP}^{\rm eff}\geq1$) or a massive hidden photon $\GP$ ($N_{\nu,\GP}^{\rm eff}=21/16$).  
The interactions of MCPs and $\GP$s with the standard bath should not allow thermalization before BBN. 
MCPs $\psi$ are produced with a rate $\Gamma(e^+ e^-\to \psi\ \overline \psi)\sim \alpha^2 Q^2_{\rm MCP} T/2$ (with $Q_{\rm MCP}$ the MCP electric charge) while $\GP$s with $\Gamma(\gamma e^\pm \to \gamma' e^\pm)\sim \chi^2_{\rm eff}\Gamma_{\rm C}$ with $\Gamma_{\rm C}$ the standard Compton scattering rate. Here $\chi_{\rm eff}$ is the effective $\G-\GP$ mixing in the plasma, which for sub-eV $\GP$ masses is $\chi_{\rm eff}\simeq \chi(\muu/\OP)^2$. The ratio of the $\GP$ mass to the plasma frequency $\muu/\OP$ is extremely small before BBN so it suppresses $\GP$ production with respect to other WISPs.
Comparing with the expansion rate $H$ we find that MCPs with $Q_{\rm MCP}< 2\times 10^{-9}$ 
would be allowed~\cite{Davidson:2000hf,Berezhiani:2008gi} but there are no significant bounds for hidden photons~\cite{Masso:2006gc,Redondo:2008tq}.

\emph{Cosmic Microwave background.-}
The today's measured CMB features an almost perfect blackbody spectrum with ${\cal O}(10^{-5})$ angular anisotropies.
It is released at $T\sim 0.1$ eV but the reactions responsible of the blackbody shape freeze out much earlier, at $T\sim$ keV. Reactions like $\gamma+...\to$WISP$+...$ will deplete photons in a frequency dependent way, which can be constrained by the precise FIRAS spectrum measurements~\cite{Fixsen:1996nj}. This has been used to constrain light MCPs~\cite{Melchiorri:2007sq} and HPs with $\muu\lesssim 0.2$ meV~\cite{Jaeckel:2008fi}.
On the other hand, around $T\sim$ eV the primordial plasma is so sparse that WISPs would free-stream out of the density fluctuations, diminishing their contrast.
Moreover, thermal WISPs contribute to the \emph{radiation} energy density, delaying the matter-radiation equality and reducing the contrast growth before decoupling. 
In these matters they act as standard neutrinos~\cite{Ichikawa:2008pz} so $\rho_x$ (and the couplings that would produce it) can again be constrained from the value of $N_\nu^{\rm eff}$ inferred from analysis of CMB anisotropies and other\footnote{One needs to complement CMB anisotropies with other LSS data to break the degeneracy of $\tilde N^{\rm eff}_{\nu,x}$ with other cosmological parameters such as the dark matter density.} large scale structure (LSS) data 
$\tilde N_{\nu,x}^{\rm eff}\equiv  \left({4/11}\right)^{4/3} N^{\rm eff}_{\nu,x} =  -0.1_{+2.0}^{-1.4}$~\cite{Simha:2008zj}. 
This argument has been used to constraint axions~\cite{Hannestad:2007dd,Hannestad:2008js} and meV $\GP$s~\cite{Jaeckel:2008fi}. In this bound Ly-$\alpha$ forest data has been deliberately omitted. 
Ly-$\alpha$ has systematically favored values of $\tilde N_{\nu,x}^{\rm eff}$ larger than zero \cite{Seljak:2006bg,Mangano:2006ur} which could be revealing the existence of a cosmic WISP relic density\footnote{Probably because of an incorrect treatment of the bias parameter~\cite{Hamann:2007pi}.}. 
If this anomaly is due to a population of $\GP$s created through resonant oscillations $\gamma-\GP$ between BBN and the CMB decoupling it can be tested in the near future by new laboratory experiments such as ALPS at DESY~\cite{Jaeckel:2008fi,Jaeckel:2007ch,Jaeckel:2008ja,Jaeckel:2008ka,Jaeckel:2008sz}.

\emph{Bounds from stellar evolution.-}
The production of WISPs in stellar interiors can substantially affect stellar evolution \cite{Raffelt:1996wa}.
WISPs can be only scarcely produced in the dense plasmas of stellar interiors, but they will easily leave the star contributing directly to its overall luminosity. 
On the other hand, only photons of the photosphere (or neutrinos) contribute to the standard energy loss. 
Therefore, the WISP luminosity is enhanced at least by a volume/surface factor and a further $(d_{\rm inside}/d_{\rm surface})^n(T_{\rm inside}/T_{\rm surface})^m$ ($d$ a relevant particle density, $n,m>1$) with respect to the standard luminosity. This can be a huge enhancement which certainly justifies the typical strong constraints.  

Stars evolve fusing increasingly heavier nuclei in their cores. Heavier nuclei require hotter environments, and when a nuclear species is exhausted in the core this slowly contracts and heatens up until it reaches the new burning phase. WISP emission shortens normal burning phases (the energy loss rate is higher than standard but the total energy is limited by the number of nuclei) but enlarges the intermediate (Red Giant) phases (WISP cooling delays reaching the appropriate temperature during the core contraction).

These effects have been used to constraint a variety of WISPs in different stellar environments~\cite{Raffelt:2006cw,Raffelt:1996wa} for which information on evolutionary time scales is  available. The strongest limits for general axion-like-particles (ALPs) with a two photon coupling and MCPs come from observations of Horizontal Branch (HB) stars in globular clusters (GC)~\cite{Raffelt:1985nk,Raffelt:1987yu}. 
For the standard QCD axions, the best constraints come from White Dwarf cooling~\cite{Raffelt:1985nj,Isern:2008nt} through the coupling to electrons (DFSV axions) and from the duration of the SN1987A neutrino burst~\cite{Raffelt:2006cw} through the nucleon coupling (KSVZ axions). 

The Sun is less sensitive than these other stars to WISP emission, even though its properties are better known. Solar bounds have been obtained from studies of its lifetime, helioseismology and the neutrino flux~\cite{Schlattl:1998fz,Gondolo:2008dd}, but although more precise they are also less demanding. Nevertheless, if WISPs are emitted from the Sun one can detect them with a dedicated laboratory experiment at earth~\cite{Sikivie:1983ip,vanBibber:1988ge,Paschos:1993yf}. 
One of the so-called Helioscope axion searches~\cite{Lazarus:1992ry,Moriyama:1998kd,Inoue:2008zp,Zioutas:2004hi}, CAST, has recently beaten the HB constraints for ALPs with a two photon coupling~\cite{Andriamonje:2007ew}, and its results have been used to limit a possible solar $\GP$ flux~\cite{Redondo:2008aa,Gninenko:2008pz}.
Following the now disclaimed~\cite{Zavattini:2007ee} PVLAS 2005 results~\cite{Zavattini:2005tm}, specific models were recently built that suppress WISP emission from stars~\cite{Masso:2005ym,Masso:2006gc,Redondo:2008tq,Brax:2007ak,Mohapatra:2006pv}. 
If this idea is realized, Helioscope bounds will gain terrain to energy loss arguments~\cite{Jaeckel:2006xm} 
 ($\GP$s are the minimal example of this case~\cite{Redondo:2008aa}).
 
In summary, cosmology and astrophysics provide the strongest constraints on the (minimal) WISP models described elsewhere in these proceedings, with the only exception of sub-meV $\GP$s. Summary plots are shown in Figs. 1 and 2. 

\begin{figure}[h]
\centerline{\includegraphics[width=.8\textwidth]{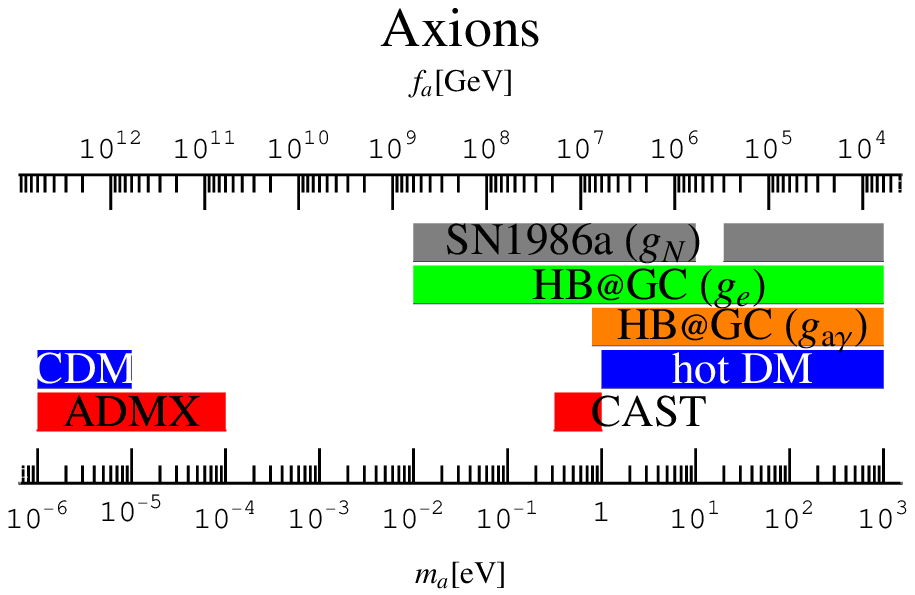}}
\vspace{0.5cm}
\centerline{\includegraphics[width=.65\textwidth]{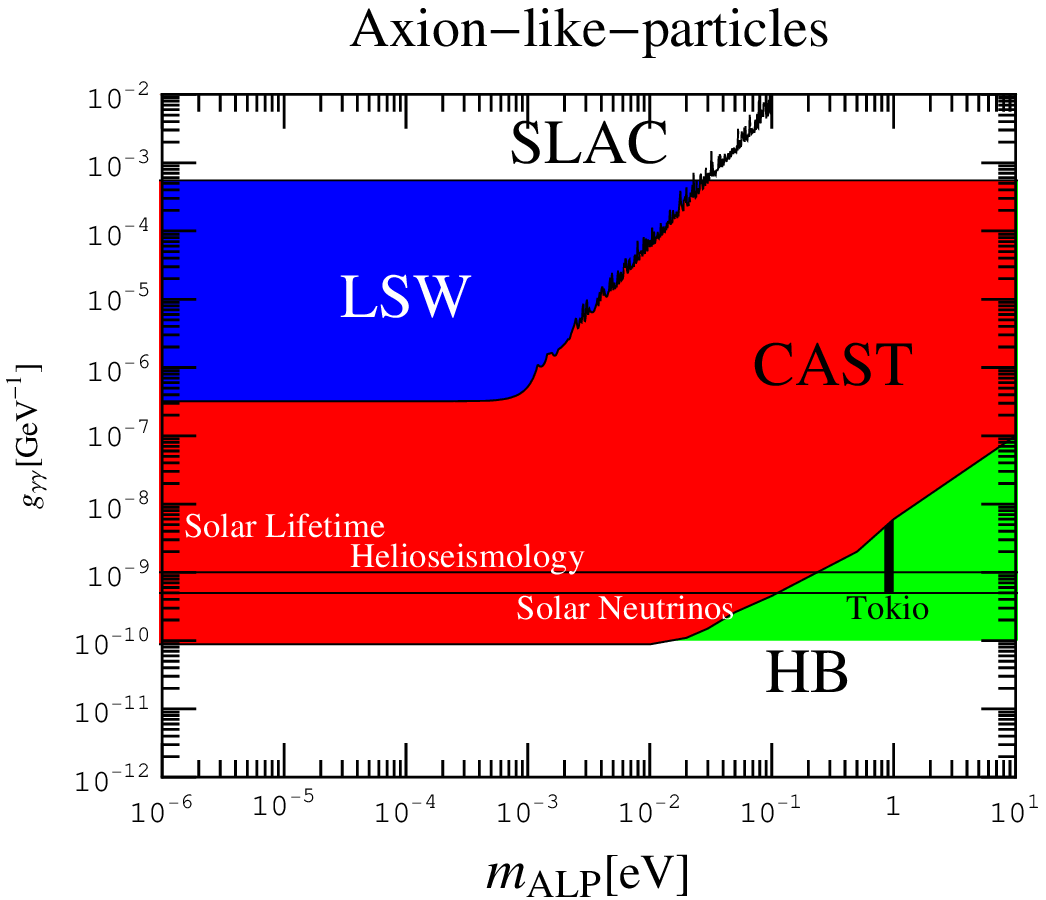}} 
\caption{\small
Summary of cosmological and astrophysical constraints for axions (up) (for the mass $m_a$ or decay constant $f_a$)~\cite{Raffelt:2006cw} and axion-like-particles (down) (two photon coupling $g_{\gamma\gamma}$ vs. mass $m_{\rm ALP}$)~\cite{Andriamonje:2007ew,Schlattl:1998fz,Inoue:2008zp}. See the text for details. For comparision, the most notable laboratory limits are also shown.}\label{Fig:MV}
\end{figure}

\begin{figure}[h]
\centerline{\includegraphics[width=.65\textwidth]{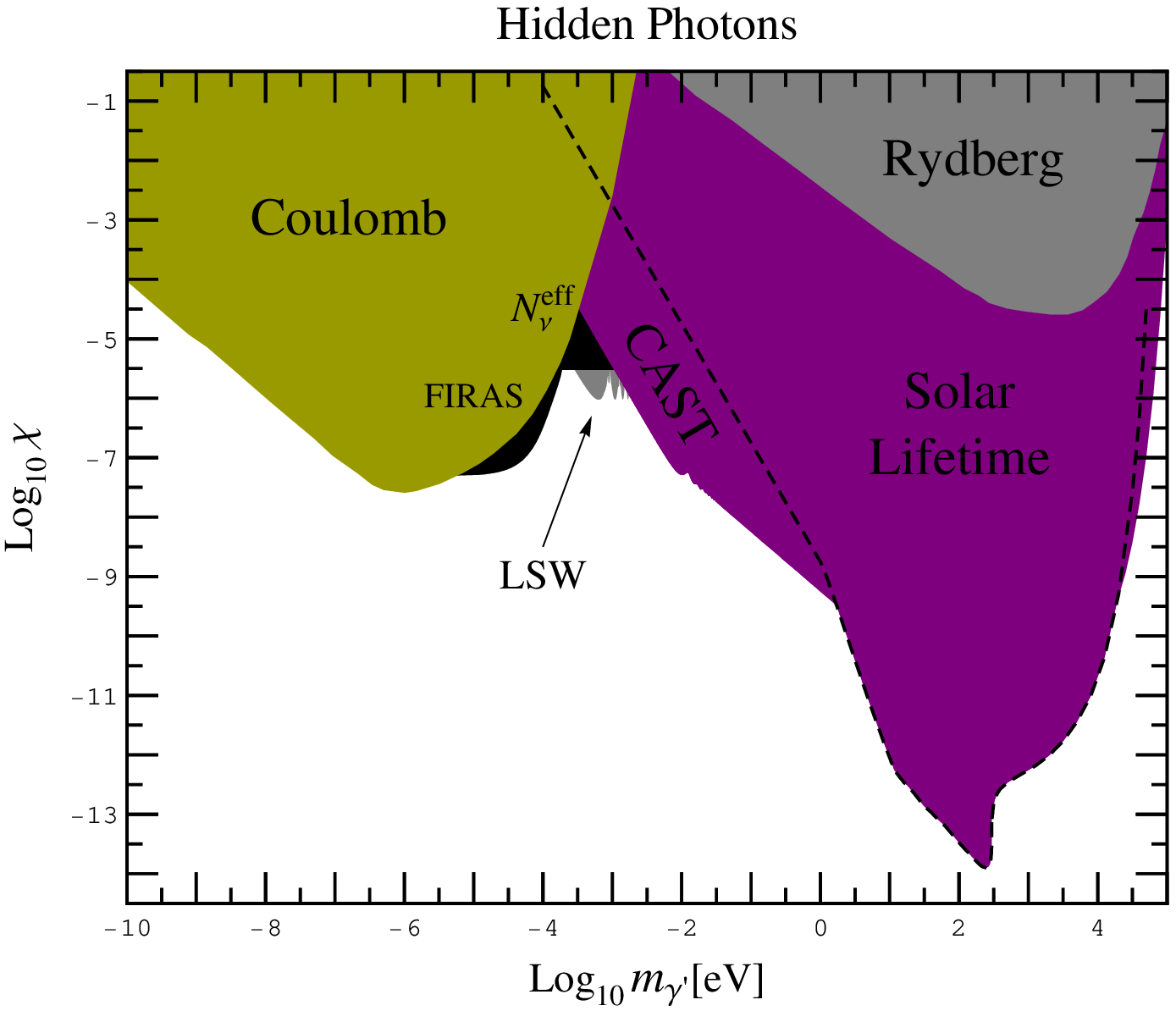}}
\vspace{0.5cm}
\centerline{\includegraphics[width=.65\textwidth]{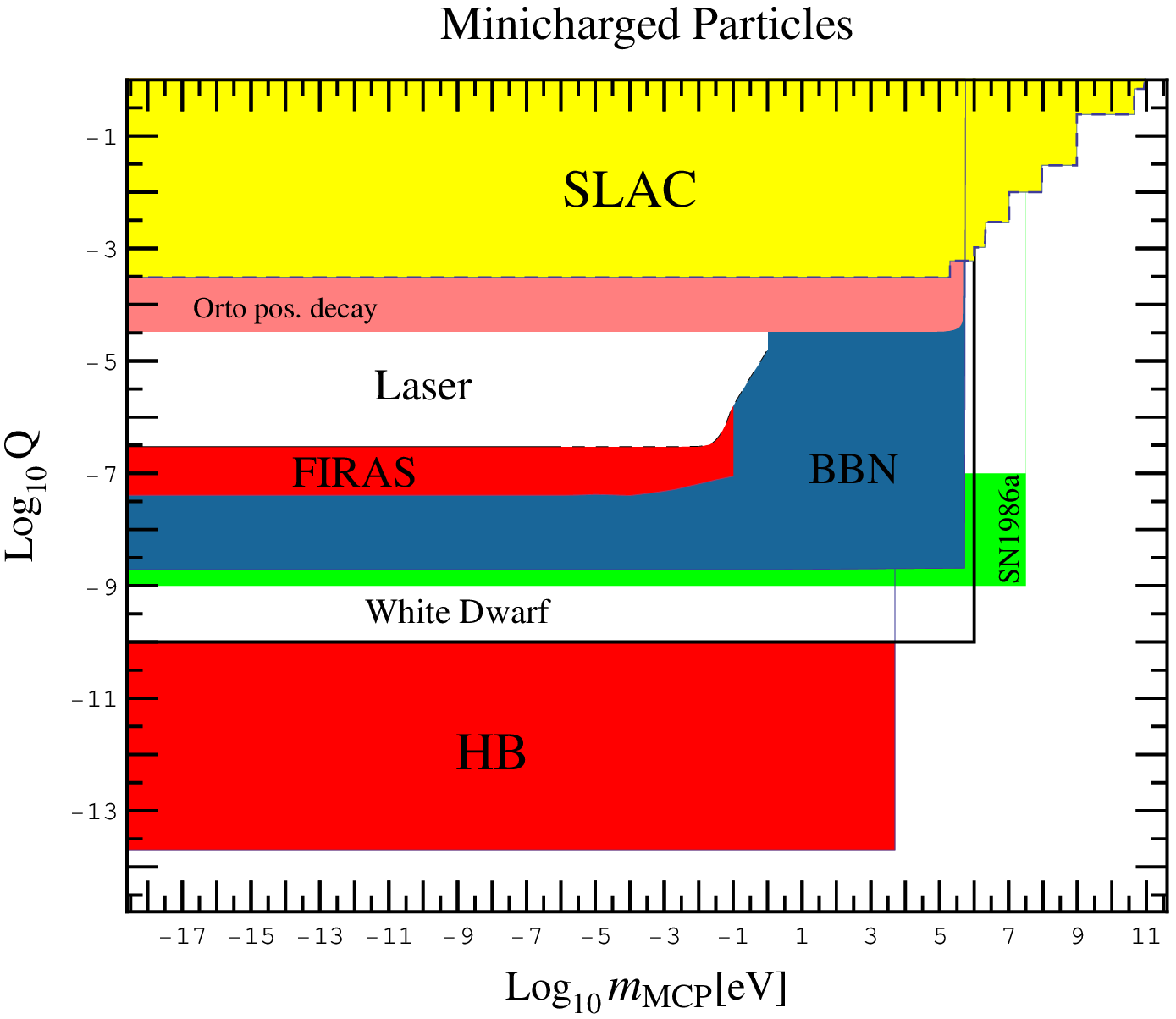}} 
\caption{\small
Summary of cosmological and astrophysical constraints for hidden photons (up) (kinetic mixing with photons $\chi$ vs. mass $\muu$)~\cite{Ahlers:2008qc} and minicharged particles (down) (charge $Q$ vd. mass $m_{\rm MCP}$)~\cite{Ahlers:2008qc}. See the text for details. For comparision, some laboratory limits are also shown.}\label{Fig:MV}
\end{figure}
\section*{Acknowledgements}
The author wishes to thank the participants of the ``4th Patras Workshop on Axions, WIMPs and
WISPs" and the ``Brainstorming and Calculationshop on the Physics Case for a Low Energy frontier" 
for very stimulating discussions in a really friendly ambient.
 
\begin{footnotesize}


\providecommand{\href}[2]{#2}\begingroup\raggedright\endgroup

\end{footnotesize}


\end{document}